\documentstyle[preprint,aps,epsf,eqsecnum,floats]{revtex}
\tighten

\begin{document}

\preprint{\vbox{\hbox{JHU--TIPAC--950008}
\hbox{hep-ph/9507311}}}

\title{Excited Heavy Mesons Beyond Leading Order in the Heavy Quark Expansion}
\author{Adam F.~Falk and Thomas~Mehen}
\address{Department of Physics and Astronomy,
The Johns Hopkins University\\
3400 North Charles Street,
Baltimore, Maryland 21218 U.S.A.\\
{\tt falk@planck.pha.jhu.edu}\\
{\tt mehen@dirac.pha.jhu.edu}}

\date{July 1995}

\maketitle
\begin{abstract}
We examine the decays of excited heavy mesons, including the leading power
corrections to the heavy quark limit.  We find a new and natural explanation
for
the large deviation of the width of the $D_1(2420)$ from the heavy quark
symmetry
prediction.  Our formalism leads to detailed predictions for the properties of
the excited bottom mesons, some of which recently have been observed.  Finally,
we present a detailed analysis of the effect of power corrections and finite
meson widths on the angular distributions which may be measured in heavy meson
decays.
\end{abstract}


\section{Introduction}

The excitation spectrum of charmed and bottom mesons has received considerable
recent theoretical and experimental attention.  The discovery of the
$B_1$ and the $B_2^*$ mesons~\cite{OPAL}, and the measurement of their masses
and
widths, complements the improving data being acquired on their charmed cousins,
the $D_1$ and $D_2^*$~\cite{CLEO}.  At the same time, the theoretical
understanding of the production and decay of these mesons has profited from the
application of the Heavy Quark Effective Theory (HQET) and the ideas behind it,
in particular the enlarged spin-flavor symmetry of QCD which obtains in the
limit
$m_c,m_b\to\infty$~\cite{HQET}.

We will begin this paper with a review of the experimental situation, and of
the implications of heavy quark symmetry for excited heavy mesons.  We will
then investigate
systematically the effect of the leading corrections to the heavy quark limit.
We focus on corrections which violate the heavy spin symmetry, since it is
predictions which follow from this symmetry which are tested by current data.
We
will find a new and natural explanation for the anomalously large width of the
$D_1$.  Our formalism leads to detailed predictions for the properties of the
$B_1$ and $B_2^*$, as well as for their strange counterparts.  We close by
examining various angular distributions in the strong decays of heavy mesons,
since with sufficiently precise data they will eventually provide detailed
tests
of our proposal.

\section{The Heavy-Light Meson Spectrum}

The excitation spectrum of heavy mesons takes a particularly simple form in the
limit $m\to\infty$, where $m=m_c$ or $m=m_b$.  In this limit, the spin of the
heavy quark decouples, and both the spin $J$ of the meson and the angular
momentum $J_\ell$ of the light degrees of freedom become good quantum numbers.
For each state in which the light degrees of
freedom have spin-parity $J_\ell^P$, there is a degenerate doublet of meson
states with $J^P=J_\ell^P\pm{1\over2}$.  The mass splitting between these
doublets arises only from effects of order $\Lambda_{\rm QCD}^2/m$ in the heavy
quark expansion.

In this language, the ground state heavy mesons $M$ and $M^*$ have light
degrees
of freedom with $J_\ell^P={1\over2}^-$, and there are low-lying excited
$P$-wave
states with $J_\ell^P={1\over2}^+$ and $J_\ell^P={3\over2}^+$.  The current
experimental situation is summarized in Table~\ref{mesondata}.  Here we quote
errors only for the excited $D$ mesons~\cite{CLEO,PDG}.  The data on the
excited
$B$ mesons is the result of a fit for which no errors are given~\cite{OPAL},
and the different charge states of the $B^*$, $B_1$ and $B_2^*$ have not been
resolved.  States for which no masses are given in Table~\ref{mesondata} have
not yet been observed, although they are expected to exist.
\begin{table}
  \begin{tabular}{cccccccc}
     \multicolumn{2}{c}{Spin}&\multicolumn{3}{c}{$D$ system~\cite{CLEO,PDG}}
           &\multicolumn{3}{c}{$B$ system~\cite{OPAL,PDG}}\\
     \hline
     $J_\ell^P$&$J^P$&state&$M$ (MeV)
     &$\Gamma$ (MeV)&state&$M$ (MeV)&$\Gamma$ (MeV)\\
     \hline\hline

${1\over2}^-$&$0^-$&$D^0$&1865&$\tau=0.42\,$ps&$B^0$&5279&$\tau=1.50\,$ps\\
     &&$D^\pm$&1869&$\tau=1.06\,$ps&$B^\pm$&5279&$\tau=1.54\,$ps\\
     &&$D_s$&1969&$\tau=0.47\,$ps&$B_s$&5375&$\tau=1.34\,$ps\\
     \cline{2-8}
     &$1^-$&$D^{*0}$&2007&$<2.1$&$B^*$&5325\\
     &&$D^{*\pm}$&2010&$<0.13$\\
     &&$D_s^*$&2110&$<4.5$&$B_s^*$\\
     \hline\hline
     ${1\over2}^+$&$0^+$&$D_0^*$&&&$B_0^*$\\
     \cline{2-8}
     &$1^+$&$D_1'$&&&$B_1'$\\
     \hline\hline
     ${3\over2}^+$&$1^+$&$D_1^0$&$2421\pm3$&$20\pm7$&$B_1$&5725&20\\
     &&$D_1^\pm$&$2425\pm3$&$26\pm9$\\
     &&$D_{s1}$&2535&$<2.3$&$B_{s1}$&5874&1\\
     \cline{2-8}
     &$2^+$&$D_2^{*0}$&$2465\pm4$&$28\pm10$&$B_2^*$&5737&25\\\
     &&$D_2^{*\pm}$&$2463\pm4$&$27\pm12$\\
     &&$D_{s2}^*$&$2573\pm2$&$16\pm6$&$B_{s2}^*$&5886&1\\
  \end{tabular}
  \caption{}
  \label{mesondata}
\end{table}

Heavy quark symmetry imposes a number of constraints on the strong decays of
these states.  Since such decays are entirely transitions of the light
degrees of freedom, the four possible decays of the two members of a doublet
with given $J_\ell^P$ to the two members of another doublet with
$J_\ell^{P\prime}$ are all essentially a single process.  In the strict
$m\to\infty$ limit, in which all
such doublets are completely degenerate, this fact leads to the simple
prediction
that the two excited states should have exactly the same width.  In fact, it is
more accurate to use the actual masses of the states in calculating phase space
effects, imposing the heavy quark symmetry only on the level of the matrix
elements.  Even though this approach is not technically consistent in the sense
of the $1/m$ expansion, it allows us to incorporate certain $1/m$ effects which
are numerically quite substantial.

For example, since the decay $D_2^*\to D\pi$ can take place only with the $\pi$
in an orbital $d$-wave, the same also must be true of the transitions $D_2^*\to
D^*\pi$ and $D_1\to D^*\pi$ (the matrix element for $D_1\to D\pi$  vanishes).
Incorporating phase space effects, the partial width for the transition $D_i\to
D_f\pi$ is proportional to $(M_{D_i}/M_{D_f})|{\bf p}_\pi|^5$, times a
mass-independent matrix element.  This fact, plus some elementary spin
counting,
allows one to make the predictions~\cite{IW}
\begin{equation}\label{widthpred}
   {\Gamma(D_2^{*0}\to D^+\pi^-)\over\Gamma(D_2^{*0}\to D^{*+}\pi^-)}=2.3\,,
   \qquad\qquad{\Gamma(D_1^0)\over\Gamma(D_2^{*0})}=0.30\,.
\end{equation}
(We choose the modes for which the data are most accurate.)  Experimentally,
one finds~\cite{CLEO,PDG}\footnote{We extract the ratio
$\Gamma(D_1^0)/\Gamma(D_2^{*0})$ from the data in Table~\ref{mesondata}.  We do
not assign an error, because the correct error depends crucially on the
correlations between the measurements of the widths, which we do not know.}
\begin{equation}
   {\Gamma(D_2^{*0}\to D^+\pi^-)\over\Gamma(D_2^{*0}\to
D^{*+}\pi^-)}=2.2\pm0.9\,,
   \qquad\qquad{\Gamma(D_1^0)\over\Gamma(D_2^{*0})}=0.71\,.
\end{equation}
Clearly, one of these predictions works extremely well, while the other works
not
at all.  Why might this be so?

One common explanation is that the $D_1$ has a small mixing with the $D_1'$,
which decays in an $s$-wave rather than a $d$-wave, and hence is expected to be
considerably broader~\cite{IW}.  Such a mixing is allowed when spin symmetry
violating $1/m$ effects are included, as the $D_1$ and $D_1'$ both transform
as $J^P=1^+$ under the Lorentz group and differ only in their values of
$J_\ell^P$.  Since even a small mixing is important if it is with a much
broader state, this provides a simple explanation of why a $1/m$ correction of
the natural size
might lead to an anomalously large correction to the total width of the $D_1$.

Unfortunately, it is not one which is particularly favored by the present data.
One may measure the angular distribution of the emitted pion and determine
directly whether it is in an $s$-wave or a $d$-wave.  The situation is
complicated by the fact that the pion angular distribution depends not
only the ratio of the $s$ and $d$-wave partial widths, but also on the
relative phase $\eta$ between the two matrix elements.  Particularly in their
data on $D_1^0$ decay~\cite{CLEO}, CLEO finds that a large $s$-wave component
is
compatible only with a restricted region in $\cos\eta$.  By no means is such a
scenario ruled out, but it is less than generic.

In addition, there is no evidence for a significant $s$-wave component in the
decay of the $D_{s1}$, which is related by flavor $SU(3)$ symmetry to the
$D_1$.
One may use heavy quark symmetry to predict the $d$-wave width of the $D_{s1}$
in
terms of the width of the $D_{s2}^*$, analogous to the second relation of
Eq.~(\ref{widthpred}).  The $D_{s1}$ and $D_{s2}^*$ decay via $K$ emission to
$D$
and $D^*$.  Given the measured $\Gamma(D_{s2}^*)$, one predicts
\begin{equation}
   \Gamma(D_{s1})=0.3\,{\rm MeV}\,,
\end{equation}
far below the CLEO upper limit.  Still, there is little room for a large
additional
$s$-wave component.  To see this, one may use flavor $SU(3)$ and the upper
limit
on $\Gamma(D_{s1})$ to predict an upper limit on $\Gamma(D_1)$~\cite{CT}.
Since
$|{\bf p}_\pi|$ in $D_1\to D^*\pi$ is typically larger than $|{\bf p}_K|$ in
$D_{s1}\to D^*K$, the upper limit on $\Gamma(D_1)$ will be much more stringent
if
the decay is assumed to be primarily $s$-wave, as opposed to $d$-wave.  One
then
obtains independent upper limits on the $s$-wave and $d$-wave components of
$\Gamma(D_1)$:
\begin{eqnarray}
   \Gamma_s(D_1)&<&3\,{\rm MeV}\,,\\
   \Gamma_d(D_1)&<&105\,{\rm MeV}\,.\nonumber
\end{eqnarray}
The correlated limits are somewhat stronger (see Ref.~\cite{CT}).  However, we
see immediately that under these assumptions it is impossible to accommodate a
large $s$-wave component in $D_1$ decay.

Of course, these assumptions might not be very good.  Flavor $SU(3)$ could fail
badly here.  One test of $SU(3)$ is to use the width of the $D_2^*$ to predict
the width of the $D_{s2}^*$, assuming $d$-wave decay and including the correct
phase space for the $K$.  One finds $\Gamma(D_{s2}^*)=(9\pm3)\,$MeV, in
reasonable agreement with experiment.  (Perhaps it would be more correct to
include the
$SU(3)$ violating factor $f_\pi^2/f_k^2$ in this prediction; however, doing so
makes the agreement with experiment worse.)  However, it is possible that
the $D_{s1}-D_{s1}'$ mixing is very different from the $D_1-D_1'$ mixing, since
the angle depends delicately on the interplay of a
mixing matrix element and a mass splitting, both of which receive
$SU(3)$-violating corrections.  Still, we are not encouraged
that this explanation for the anomalously large $D_1$ width is the correct one.

As an alternative, it has been suggested~\cite{EHQ} that the $D_1$ width
receives
a large contribution from the emission of two pions which resonate through a
$\rho$ meson, with no analogous enhancement in $D_2^*$ decay.  Two pion
decays in which one of the pions resonates with a broad $D_1'$ or $D_0^*$ have
also been considered~\cite{FL}, but are not thought to contribute significantly
to the total width.  In the next section, we will introduce a simpler
explanation
for the $D_1$ width, which arises naturally at higher order in the heavy quark
expansion.

\section{The Chiral Lagrangian}

The strong decays of excited mesons involve the emission of soft pions and
kaons, and hence it is useful to analyze these interactions with the help of
chiral perturbation theory.  To be concrete, we will specify to the charm
system; the generalization to bottom is at all points straightforward.
The chiral lagrangian appropriate to the analysis of ground state and excited
heavy mesons has been derived elsewhere~\cite{HHCPT}; here we will simply
recall
the basic points.

The heavy mesons are represented by matrix superfields which carry a
representation
not only of the Lorentz group but also of the $SU(2)$ heavy quark spin
symmetry.  In addition, they transform as $\overline 3$'s under flavor $SU(3)$,
since by convention our heavy mesons contain a single heavy quark (rather than
an antiquark).  For the ground state and lowest excited states discussed in the
previous section, these superfields take the form~\cite{FGGW}
\begin{eqnarray}
  H_a &=& {(1+\rlap/v)\over2\sqrt2}
  \left[ D_a^{*\mu}\gamma_\mu-D_a\gamma^5\right]\,,\nonumber\\
  S_a &=& {(1+\rlap/v)\over2\sqrt2}
  \left[ D_{1a}^{\prime\mu}\gamma_\mu\gamma^5-D_{0a}^*\right]\,,\\
  T^\mu_a &=& {(1+\rlap/v)\over2\sqrt2}\left[ D_{2a}^{*\mu\nu}\gamma_\nu
  -D_{1a}^\nu\sqrt{\case3/2}\,\gamma^5(\delta^\mu_\nu-\textstyle{1\over3}
  \gamma_\nu\gamma^\mu+\textstyle{1\over3}\gamma_\nu v^\mu)\right]\,,\nonumber
\end{eqnarray}
where $v^\mu$ is the four-velocity of the heavy meson, and $a$ is the flavor
index.  These fields are normalized nonrelativistically.  Under a heavy quark
spin rotation $S_Q$, the superfields transform as $H\to S_QH$, etc., while
under a Lorentz transformation $S_L$ they transform as $H\to S_LHS_L^\dagger$,
etc..

The heavy mesons interact with the octet of pseudo-Goldstone bosons, which are
treated with the usual nonlinear formalism.  The lagrangian is written in terms
of an exponentiated matrix of boson fields, $\xi=\exp(i{\cal M}/f_\pi)$, where
\begin{equation}
   {\cal M}=\pmatrix{
   \sqrt{1\over2}\ \pi^0+\sqrt{1\over6}\ \eta & \pi^+ & K^+ \cr
   \pi^- & -\sqrt{1\over2}\ \pi^0+\sqrt{1\over6}\ \eta & K^0 \cr
   K^- & \overline K^0 & -\sqrt{2\over3}\ \eta \cr}
\end{equation}
and $f_\pi\approx135\,{\rm MeV}$.  Under $SU(3)_L\times SU(3)_R$, the field
$\xi$ transforms as $\xi\to L\xi U^\dagger=U\xi R^\dagger$, where $U$ is a
matrix
which depends on ${\cal M}$.  For generators within the diagonal subgroup,
$U=L=R$.

The pseudo-Goldstone bosons couple to the heavy superfields through the
covariant
derivative $D^\mu_{ab}=\delta_{ab}\partial^\mu+{1\over2}
(\xi^\dagger\partial^\mu\xi + \xi\partial^\mu\xi^\dagger)_{ab}$ and the axial
vector field $A^\mu_{ab}={i\over2}(\xi^\dagger\partial^\mu\xi -
\xi\partial^\mu\xi^\dagger)_{ab}$.  Under $SU(3)_L\times SU(3)_R$, the fields
transform as $A\to UAU^\dagger$, $H\to UH$, $D_\mu H\to UD_\mu H$, and so on.
The kinetic part of the chiral lagrangian then given by
\begin{eqnarray}
   {\cal L}_{\rm kin} &=& {1\over8}f_\pi^2\,{\rm Tr}\,\left[
   \partial^\mu\Sigma\partial_\mu\Sigma^\dagger\right]
   - {\rm Tr}\,\left[\overline Hiv\cdot DH\right]
   + {\rm Tr}\,\left[\overline S(iv\cdot D-\Delta_S)S\right]\nonumber\\
   &&\qquad\mbox{}
   + {\rm Tr}\,\left[\overline T^\mu(iv\cdot D-\Delta_T)T_\mu\right]\,,
\end{eqnarray}
where $\Sigma=\xi^2$.  The trace is taken with respect to spinor and flavor
indices, which we suppress.  The excitation energies $\Delta_S$ and $\Delta_T$
are defined with respect to the spin-averaged masses of the excited doublets.
The heavy fields obey the equations of motion $iv\cdot DH=0$, $iv\cdot
DS=\Delta_SS$ and $iv\cdot DT^\mu=\Delta_TT^\mu$.

The interactions between the various fields are constrained by heavy spin
symmetry and chiral symmetry,\footnote{Since we are only interested in
transitions involving a single light meson, flavor $SU(3)$, under which the
pions transform linearly, would be sufficient for our analysis.  We employ the
full formalism of nonlinear representations simply as a convenience.} as well
as
by Lorentz invariance (including parity and time reversal) and velocity
reparameterization invariance~\cite{CT,FL,LM}.  We are interested in matrix
elements between the ground state doublet and each of the excited states, for
which we introduce the interaction terms
\begin{eqnarray}
   {\cal L}_s&=&f\,{\rm Tr}\,\left[\overline H\,
   S\gamma^\mu\gamma^5A_\mu\right]
   +{\rm h.c.}\,,\nonumber\\
   {\cal L}_d&=&{h\over\Lambda_\chi}\,{\rm Tr}\,\left[\overline H\,
   T^\mu\gamma^\nu\gamma^5
   (iD_\mu A_\nu+iD_\nu A_\mu)\right]+{\rm h.c.}\,.
\end{eqnarray}
These are the leading terms in the chiral expansion, which is an expansion in
derivatives and fields over the chiral symmetry breaking scale
$\Lambda_\chi\sim1\,{\rm GeV}$.  Note that ${\cal L}_d$, which mediates the
$d$-wave decays of the $D_1$ and $D_2^*$, appears at higher order than ${\cal
L}_s$, which mediates the $s$-wave decays of the $D_0$ and $D_1'$.  Velocity
reparameterization invariance plays a particularly important role in
constraining ${\cal L}_d$~\cite{CT}.

One may include bottom mesons by introducing new superfields $H_b$, $S_b$ and
$T^\mu_b$ to represent the ground state and excited $B$ mesons.  These fields
appear in analogues of the terms ${\cal L}_{\rm kin}$, ${\cal L}_s$ and ${\cal
L}_d$, with the {\it same\/} parameters $\Delta_S$, $\Delta_T$, $f$ and $h$ as
in
the case of charm.  This doubling of terms persists at higher orders, again
with
identical coefficients, except that explicit factors of $1/m$ should be taken
to
be $1/m_c$ or $1/m_b$ as appropriate.

The chiral lagrangian, as developed so far, is sufficient to make the heavy
spin
symmetry predictions
discussed in Section II.  However, as we have seen, certain of these
predictions
work better than others; in particular, the ratio of partial widths
$\Gamma(D_2^*\to D^*\pi)/\Gamma(D_2^*\to D\pi)$ is well predicted by the heavy
spin symmetry, while the ratio of full widths $\Gamma(D_1)/\Gamma(D_2^*)$ is
not.  Evidently, corrections to the heavy quark limit can be quite large.  Why
is
this so in some cases, but not in others?  To gain insight into this question,
it
is necessary to go to subleading order in the $1/m$ expansion.

The most general extension of the chiral lagrangian to order $1/m$ is extremely
unwieldy.  However, we are interested only in those $1/m$
corrections which break the heavy spin symmetry.  We will find
that when we neglect systematically spin symmetry conserving terms, the
resulting theory is sufficiently constrained to yield
interesting information.

The $1/m$ corrections to the chiral lagrangian arise from $1/m$ corrections to
the heavy quark lagrangian, which is derived from the full QCD lagrangian in
the $m\to\infty$ limit.  This expansion takes the form~\cite{FGL}
\begin{equation}
   {\cal L}_{\rm HQET} = \bar hiv\cdot Dh+{1\over2m}\bar h(iD)^2h
   +{1\over2m}\bar h\sigma^{\mu\nu}(\textstyle{1\over2}gG_{\mu\nu})h
   +\dots\,,
\end{equation}
where $h$ is the HQET heavy quark field.  The effect of the subleading terms
$\bar h(iD)^2h$ and $g\bar h\sigma^{\mu\nu}G_{\mu\nu}h$ on the chiral expansion
may be treated in the same manner as other symmetry breaking perturbations to
the fundamental theory such as finite light quark masses.  Namely, we
introduce a ``spurion'' field which carries the same representation of the
symmetry group as does the perturbation in the fundamental theory, and then
include this spurion in the chiral lagrangian in the most general
symmetry-conserving way.  When the spurion is set to the constant value which
it
has in QCD, the symmetry breaking is transmitted to the effective theory.  In
the
case of finite light quark masses, for example, the symmetry breaking term in
QCD
is $\bar q M_qq$, where $M_q={\rm diag}(m_u,m_d,m_s)$.  Introducing a spurion
$M_q$ which transforms as $M_q\to LM_qR^\dagger$ under chiral $SU(3)$, we
then include terms in the ordinary chiral lagrangian such as
$\mu\,{\rm Tr}\,[M_q\Sigma+M_q\Sigma^\dagger]$.

In the present case, only the second of the two correction terms in ${\cal
L}_{\rm HQET}$ violates the heavy spin symmetry.  We include its effect in the
chiral lagrangian by introducing a spurion $\Phi_s^{\mu\nu}$ which transforms
as
$\Phi_s^{\mu\nu}\to S_Q\Phi_s^{\mu\nu}S_Q^\dagger$ under a heavy quark spin
rotation $S_Q$.  This spurion is introduced in the most general manner
consistent with heavy quark symmetry, and is then set to the constant
$\Phi_s^{\mu\nu}={1\over2m}\sigma^{\mu\nu}$ to yield the leading spin symmetry
violating corrections to the chiral lagrangian.  We will restrict ourselves to
terms in which $\Phi_s^{\mu\nu}$ appears exactly once.

The simplest spin symmetry violating effect is to break the degeneracy of the
heavy meson doublets.  This occurs through the terms
\begin{equation}
   \lambda_H\,{\rm Tr}\,\left[\overline H\Phi_s^{\mu\nu}H\sigma_{\mu\nu}\right]
   -\lambda_S\,{\rm Tr}\,\left[\overline S\Phi_s^{\mu\nu}S
   \sigma_{\mu\nu}\right]
   -\lambda_T\,{\rm Tr}\,\left[\overline T^\alpha\Phi_s^{\mu\nu}T_\alpha
   \sigma_{\mu\nu}\right]\,.
\end{equation}
The dimensionful coefficients are fixed once the masses of the mesons are
known.  For the ground state $D$ and $D^*$, for example, we find
\begin{equation}
   \lambda_H={1\over8}\left[M^2_{D^*}-M^2_D\right]=(260\,{\rm MeV})^2\,.
\end{equation}
This value is entirely consistent with what one would obtain, instead, with the
$B$ and $B^*$ mesons.  For the $D_1$ and $D_2^*$, we find
\begin{equation}
   \lambda_T={3\over16}\left[M^2_{D_2^*}-M^2_{D_1}\right]=(190\,{\rm MeV})^2\,.
\end{equation}
Note that $\sqrt{\lambda_H}$ and $\sqrt{\lambda_T}$ are of order hundreds of
MeV,
the scale of the strong interactions.

We are interested in the spin symmetry violating corrections to transitions in
the class $T^\mu\to H\pi$, which will arise from terms analogous to ${\cal
L}_d$
but with one occurrence of $\Phi_s^{\mu\nu}$.  The spin symmetry, along with
the
symmetries which constrained ${\cal L}_d$, requires that any such term be of
the
generic form
\begin{equation}
   {1\over\Lambda_\chi}{\rm Tr}\,\left[\overline H\Phi_s^{\mu\nu}T^\alpha
   C_{\mu\nu\alpha\beta\kappa}\gamma^5
   \left(iD^\beta A^\kappa+iD^\kappa A^\beta\right)\right]+{\rm h.c.}\,,
\end{equation}
where $C_{\mu\nu\alpha\beta\kappa}$ is an arbitrary product of Dirac matrices
and may depend on the four-velocity $v^\lambda$.  This would seem to allow for
a
lot of freedom, but it turns out that there is only a {\it single\/} term which
respects both parity and time reversal invariance:
\begin{equation}
   {\cal L}_{d1} = {h_1\over2m\Lambda_\chi}{\rm Tr}\,
   \left[\overline H\sigma^{\mu\nu}T^\alpha
   \sigma_{\mu\nu}\gamma^\kappa\gamma^5
   \left(iD_\alpha A_\kappa+iD_\kappa A_\alpha\right)\right]+{\rm h.c.}\,.
\end{equation}
We expect the new coefficient $h_1$, which has mass dimension one, to be of
order hundreds of MeV.

Finally, there may be additional correction terms which come about by
the application of velocity reparameterization invariance (VRI)~\cite{LM} to
the
leading interaction term ${\cal L}_d$.  This is a ``symmetry''\footnote{We use
quotes because this is not a physical symmetry of nature, i.e. there is no
associated conserved charge.} which, as does a gauge symmetry, arises because
of
a redundancy in the variables appearing in the lagrangian.  The four-velocity
$v^\lambda$ which describes
the heavy meson field is arbitrary up to terms of order $1/m$, and the
lagrangian must be constructed so as to be invariant under reparameterizations
of the form $v^\lambda\to v^\lambda+\epsilon^\lambda/m$, where the components
of
$\epsilon^\lambda$ are of order $\Lambda_{\rm QCD}$.  The heavy meson fields
also transform nontrivially under velocity reparameterization.  VRI is a
symmetry which constrains the new terms which may appear at higher order in the
$1/m$ expansion in terms of those which are already there.  The terms which VRI
generates at order $1/m^{n+1}$ may be found by making the
replacements~\cite{CT,LM}
\begin{eqnarray}\label{VRI}
   v^\lambda&\to&v^\lambda+{1\over m}iD^\lambda\,,\nonumber\\
   H&\to&H+{1\over2m}\left[\gamma^\nu,iD_\nu H\right]\,,\nonumber\\
   S&\to&S+{1\over2m}\left\{\gamma^\nu,iD_\nu S\right\}\,,\\
   T^\mu&\to&T^\mu+{1\over2m}\left[\gamma^\nu,iD_\nu T^\mu\right]
   -{1\over m}v^\mu iD_\nu T^\nu\nonumber
\end{eqnarray}
in the lagrangian at order $1/m$.  By the same token, all terms at order
$1/m^{n+1}$ with derivatives acting on heavy meson fields must be consistent
with such replacements at one order lower.

New interaction terms of the same order as ${\cal L}_{d1}$ will be generated
when we make the replacements~(\ref{VRI}) in the leading term ${\cal L}_d$.
However, massaging the new terms with integration by parts and application of
the equations of motion, we find that they all may be written in the form ${\rm
Tr}\left[\overline H\,T^\mu f(\partial,A)\right]$, for some Dirac matrix valued
function $f(\partial,A)$.  Hence they do not break the heavy spin symmetry and
we may ignore them for our analysis.

The other potentially important effect of $1/m$ corrections on the decay of the
$D_1$ is a possible mixing between this state and the $D_1'$.  The $D_1'$
transforms the same as the $D_1$ under the Lorentz group, but differently under
the heavy spin symmetry, so spin symmetry violating effects can mix the two
states.  Such a mixing can have a dramatic effect on the
width of the $D_1$, which now may decay via $s$-wave pion emission.

The following parity and time reversal invariant term will induce a nonzero
matrix element for the mixing of the $D_1$ and $D_1'$:
\begin{equation}\label{Lmix}
   {\cal L}_{\rm mix} = g_1{\rm Tr}\,\left[\overline S\Phi_s^{\mu\nu}T_\mu
   \sigma_{\nu\alpha}v^\alpha\right]+{\rm h.c.}\,.
\end{equation}
Unfortunately, the magnitude of the mixing matrix element does not by itself
determine the mixing angle $\psi$.  To find $\psi$, we also need the lowest
order splitting $\Delta_T-\Delta_S$ between the masses of the two states.
Then the mixing angle is given by
\begin{equation}\label{tanpsi}
   \tan\psi = {\sqrt{\delta^2+\delta_g^2}-\delta\over\delta_g}\,,
\end{equation}
where $\delta=\case1/2(\Delta_T-\Delta_S)$ and
$\delta_g=-\sqrt{2\over3}\,g_1/m$.
Since the $D_1'$ has not yet been observed,\footnote{It is possible that the
$D_J(2440)^\pm$, reported
by the TPC Collaboration~\cite{TPC} in the $D^{*0}\pi^\pm$ channel with a width
of $40\,$MeV, is the $D_1'^\pm$.  However, this observation has not been
confirmed.} we do not know $\Delta_S$; hence it is more convenient to
treat $\psi$ itself as a free parameter, rather than write it as in
Eq.~(\ref{tanpsi}).

\section{Experimental Implications}

\subsection{The $D_1$ and $D_2^*$ widths}

The new interaction term ${\cal L}_{d1}$ affects the decays of $D_1$ and
$D_2^*$
in ways that do not necessarily respect the heavy spin symmetry predictions
(\ref{widthpred}).  It is straightforward to compute the single pion partial
widths of these excited states in terms of the coupling constants $h$ and
$h_1$.  Lorentz
invariance requires that the decays of the $D_2^*$ still involve the emission
of
a $d$-wave pion, and we find
\begin{eqnarray}\label{d2widths}
   \Gamma(D_2^{*0}\to D\pi)&=&{1\over10\pi}\,{M_D\over M_{D_2^*}}
   \,{4|{\bf p}_\pi|^5\over
   \Lambda_\chi^2f_\pi^2}\left[h-{h_1\over m_c}\right]^2\,,\nonumber\\
   \Gamma(D_2^{0*}\to D^*\pi)&=&{3\over20\pi}\,{M_{D^*}\over M_{D_2^*}}
   \,{4|{\bf p}_\pi|^5\over
   \Lambda_\chi^2f_\pi^2}\left[h-{h_1\over m_c}\right]^2\,.
\end{eqnarray}
Here and in Eq.~(\ref{d1width}) below we include emission of both charged and
neutral pions, neglecting the small phase space differences between the two
channels. Note that the heavy
quark symmetry prediction for the ratio of the $D_2^*$ partial widths is
unaffected by the correction ${\cal L}_{d1}$.  This is good, because we have
seen
that the lowest order prediction works quite well already.  Truncating
Eq.~(\ref{d2widths}) at lowest order, inserting the experimental width of the
$D_2^*$, and taking $\Lambda_\chi=1\,$GeV, we obtain the estimate
$h\approx0.3$.

The effect of ${\cal L}_{d1}$ on $D_1$ decay is more complicated, because the
new
term can mediate both $d$-wave and $s$-wave decays.  The decay width is given
by
\begin{equation}\label{d1width}
   \Gamma(D_1\to D^*\pi)={1\over4\pi}\,{M_{D^*}\over M_{D_1}}
   \,{4|{\bf p}_\pi|^5\over
   \Lambda_\chi^2f_\pi^2}\left[\left(h+{5h_1\over 3m_c}\right)^2
   +{8h_1^2\over9m_c^2}\right]\,.
\end{equation}
The first term corresponds to a $d$-wave pion and the second to an $s$-wave
pion.  Note that here the $s$-wave width is also suppressed by $|{\bf
p}_\pi|^5$,
and is in no sense intrinsically larger than the $d$-wave width.  It is
consistent to neglect it at this order, since we have not included $1/m^2$
interaction terms in the lagrangian.  The corrected ratio of the $D_2^*$ width
to
the $D_1$ width is then given by
\begin{equation}\label{ratiopred0}
   {\Gamma(D_1^0)\over\Gamma(D_2^{*0})}=0.30\cdot\left[1+{16\over3}
   {h_1\over m_ch}
   +{\cal O}\left({1\over m_c^2}\right)\right]\,,
\end{equation}
where we assume that the widths are saturated by the one pion decays.
We see that in this prediction, the correction is enhanced by a large numerical
prefactor.  If we generalize Eq.~(\ref{ratiopred0}) to include a mixture of
$d$-wave and $s$-wave pion emission, we find
\begin{equation}\label{ratiopred}
   {\Gamma(D_1)\over\Gamma(D_2^*)}=0.30\left[\cos^2\psi
   \left(1+{16h_1\over3m_ch}
   \right)+\sin^2\psi{3f^2\over4(h/\Lambda_\chi)^2}\,{E_\pi^2\over|{\bf
   p}_\pi|^4}\right]\,.
\end{equation}
With $\Lambda_\chi=1\,$GeV, $h=0.3$, and $m_c=1.5\,$GeV, this reduces to
\begin{equation}\label{constraint}
   {\Gamma(D_1)\over\Gamma(D_2^*)}=0.30\left[\left(1+
   {h_1\over85\,{\rm MeV}}\right)\cos^2\psi+77f^2\sin^2\psi\right]
   {\buildrel{\rm exp.}\over=}\;0.71\,.
\end{equation}
We may saturate the experimental width of the $D_1$ with $d$-wave decay by
taking $h_1=115\,$MeV, which is quite reasonably small in view of the sizes of
the similar corrections $\lambda_H$ and $\lambda_T$.  Alternatively, if
$h_1=0$, then $(77f^2-1)\sin^2\psi=1.37$, or $f\sin\psi\approx0.13$ if $f$ is
of order one.

\subsection{Predictions for excited $B$ mesons}

The heavy spin symmetry may also be combined with the heavy flavor symmetry to
predict the masses and widths of the $B$ mesons in terms of those in the charm
sector.  The splittings between excited doublets and the ground state should be
independent of heavy quark mass, while spin symmetry violating intradoublet
splittings scale like $1/m$.  If we define the spin-averaged masses and mass
splittings
\begin{eqnarray}
   \overline M_{B}&=&\case34 M_{B^*}+\case14 M_{B}\,,\nonumber\\
   \overline M_{B^{**}}&=&\case58 M_{B_2^*}+\case38 M_{B_1}\,,\\
   \Delta M_{B^{**}}&=& M_{B_2^*}- M_{B_1}\nonumber\,,
\end{eqnarray}
and analogously for charm, the heavy quark symmetries predict
\begin{eqnarray}
   \overline M_{B^{**}}-\overline M_{B}&=&
   \overline M_{D^{**}}-\overline M_{D}\,,\nonumber\\
   \Delta M_{B^{**}}&=& {m_c\over m_b}\,\Delta M_{D^{**}}\,.
\end{eqnarray}
With $m_c/m_b=1/3$, and averaging over the charged and neutral charmed mesons,
these relations yield
\begin{eqnarray}\label{Bmasses}
   M_{B_1}&=&5780\,{\rm MeV}\,,\nonumber\\
   M_{B_2^*}&=&5794\,{\rm MeV}\,,\nonumber\\
   M_{B_{s1}}&=&5886\,{\rm MeV}\,,\\
   M_{B_{s2}^*}&=&5899\,{\rm MeV}\,.\nonumber
\end{eqnarray}
Compared with the data in Table~\ref{mesondata}, we see that the measured
masses
are somewhat lower than expected, especially for the nonstrange mesons.  The
leading corrections to the predictions for $\overline M_{B^{**}}$ and
$\overline M_{B_s^{**}}$ are of order
\begin{equation}
   \delta\sim\Lambda_{\rm QCD}^2\left({1\over2m_c}-{1\over2m_b}\right)\sim
   40\,{\rm MeV}\,,
\end{equation}
where we have estimated a QCD scale $\Lambda_{\rm QCD}\sim400\,$MeV, so the
accuracy with which the predictions (\ref{Bmasses}) work is more or less what
one would expect.

It is somewhat more delicate to make predictions for the widths of the excited
$B$ mesons, since these depend on the available phase space, hence on the
values of the heavy meson masses.  For the $B_1$ and $B_{s1}$, they also depend
on what one assumes about $J_\ell^P=\case3/2^+$ and $\case1/2^+$ mixing in the
bottom sector.  Let us introduce a notation for the pion momenta which arise in
these decays:
\begin{eqnarray}
   B_2^*\to B\pi:&&\qquad {\bf p}_{2B}\,,\nonumber\\
   B_2^*\to B^*\pi:&&\qquad {\bf p}_{2B^*}\,,\nonumber\\
   B_1\to B^*\pi:&&\qquad {\bf p}_{1B^*}\,,\nonumber
\end{eqnarray}
and similarly for charm.  Then, assuming the dominance of the one pion decay
channel, the width of the $B_2^*$ is related to that of the $D_2^*$ via
\begin{equation}\label{B2formula}
   {\Gamma(B_2^*)\over\Gamma(D_2^*)}={M_{D_2^*}\over M_{B_2^*}}\,
   \left\{{0.4|{\bf p}_{2B}|^5 M_{B} + 0.6|{\bf p}_{2B^*}|^5 M_{B^*}\over
   0.4|{\bf p}_{2D}|^5 M_{D} + 0.6|{\bf p}_{2D^*}|^5 M_{D^*}}\right\}\,,
\end{equation}
from which we find the prediction
\begin{equation}
   \Gamma(B_2^*)=(16\pm6)\,{\rm MeV}\,,
\end{equation}
with the masses given in Table~\ref{mesondata}.  This width is somewhat low,
but it is extremely sensitive to the mass of the $B_2^*$ and grows rapidly with
$M_{B_2^*}$.  Perhaps here we have a hint that when the masses of the $B_1$ and
$B_2^*$ are better measured, they will be closer to the values predicted by
heavy quark symmetry.

We may also generalize Eq.~(\ref{ratiopred}) to predict the ratio of the widths
of the $B_1$ and the $B_2^*$, assuming once again that one pion decays
dominate.  Leaving the dependence on the meson masses explicit, we find
\begin{eqnarray}\label{B1formula}
   {\Gamma(B_1)\over\Gamma(B_2^*)}&=&{M_{B_2^*}\over M_{B_1}}\,
   \left[{|{\bf p}_{1B^*}|^5 M_{B^*}\over
   0.4|{\bf p}_{2B}|^5 M_B+0.6|{\bf p}_{2B^*}|^5 M_{B^*}}\right]\nonumber\\
   &&\qquad\times\left\{\cos^2\psi_b\left(1+{16h_1\over3m_bh}\right)
   +\sin^2\psi_b{3f^2\over4(h/\Lambda_\chi)^2}\,
   {E_{1B^*}^2\over|{\bf p}_{1B^*}|^4}\right\}\,.
\end{eqnarray}
Since the mixing is generated by the spin symmetry violating operator ${\cal
L}_{\rm mix}$~(\ref{Lmix}), it should scale inversely with the heavy quark
mass.  Hence
for small mixing angles, we might expect $\psi_b\approx(m_c/m_b)\psi$.
However, as we see in Eq.~(\ref{tanpsi}), $\psi_b$ also depends delicately on
the mass splitting between the $B_1$ and the $B_1'$, so we should put no
particular trust in this estimate of $\psi_b$.  Instead, we will make specific
predictions of $\Gamma(B_1)$ only in the two
limits of pure $d$-wave and pure $s$-wave decays of {\it both\/} the $D_1$ and
the $B_1$, using the masses in Table~\ref{mesondata}:
\begin{eqnarray}
   \Gamma(B_1)/\Gamma(B_2^*)&=&0.9\qquad\qquad\hbox{pure
$d$-wave}\,,\nonumber\\
   \Gamma(B_1)/\Gamma(B_2^*)&=&1.4\qquad\qquad\hbox{pure $s$-wave}\,.
\end{eqnarray}
For the $s$-wave case, we take  $\psi_b=\psi=\pi/2$ and choose $f$ to give the
correct $D_1$ width.  While this extreme limit is not favored by the data on
$D_1$
decay, it yields a useful upper bound on $\Gamma(B_1)$.  We see that $d$-wave
dominance is somewhat favored by the current data, from which one finds
$\Gamma(B_1)/\Gamma(B_2^*)\approx0.8$.

Finally, we make predictions for the one kaon widths of the excited strange $B$
mesons.  Applying the analogue of Eq.~(\ref{B2formula}), we find
\begin{equation}
    \Gamma(B_{s2}^*)=(7\pm3)\,{\rm MeV}
\end{equation}
and from the analogue of Eq.~(\ref{B1formula}),
\begin{equation}
    \Gamma(B_{s1})/\Gamma(B_{s2}^*)=0.4\,,
\end{equation}
where the $B_{s1}$ is assumed to decay via the emission of a $K$ in a $d$-wave.
 These predictions will be tested as the data on the $B_{s2}^*$ and $B_{s1}$
improve.

\subsection{Angular distributions}

While the explanation of the width of the $D_1$ which we have presented is
certainly consistent with the data, we would like to be able to test it in
somewhat more detail.  We may do so by considering the angular distributions of
the pions emitted in its decay.  These distributions will depend on the chiral
lagrangian parameters $h$, $h_1$ and $f$, and on the mixing angle $\psi$.
Finally, with
our phase conventions the coefficient of ${\cal L}_s$ may be complex, so we
will
take $f\to f\exp(i\eta)$.

In addition to the two constraints on the set of parameters
$\{h,h_1,f,\eta,\psi\}$ from the experimental widths of the $D_2^*$ and the
$D_1$,
we will now assume that the mixing angle $\psi$ is small.  Hence we will drop
terms which are suppressed by $\sin\psi$, unless they are enhanced by a large
phase space factor.  Note that for values of $f$ of order unity, the constraint
(\ref{constraint}) requires that $\psi$ not be too large.  An estimate of
$\psi$ based on a quark wavefunction model gives
$\psi\approx9^\circ$~\cite{GIK}, which perhaps also supports the use of this
approximation.

\subsubsection{Two-pion distributions}

\begin{figure}
\epsfxsize=16cm
\hfil\epsfbox{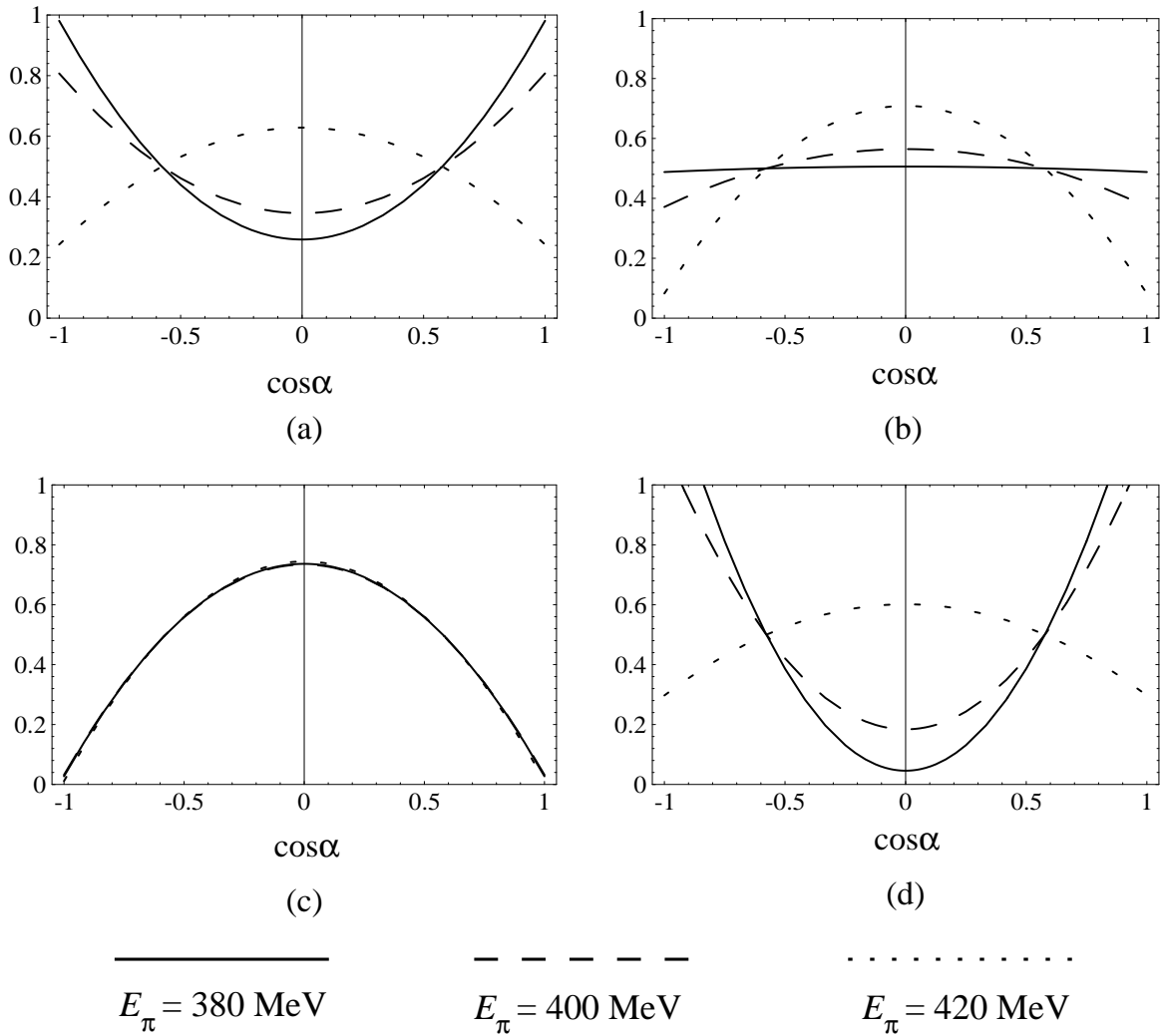}\hfill
\caption{The differential distribution in $\cos\alpha$, for the $D_1$ decaying
(a) in a pure $d$-wave; (b) in a pure $s$-wave; (c) in a mixture with $h_1$=0,
$\psi=9^\circ$ and $f=-0.85$; (d) in a mixture with $h_1$=0, $\psi=9^\circ$ and
$f=0.85$.  The solid curve is for $E_\pi=380\,$MeV, the long dashed curve for
$E_\pi=400\,$MeV, and the short dashed curve for $E_\pi=420\,$MeV.  We have
included the finite widths of the $D_1$ and $D_2^*$, and have set $\eta=0$.}
\label{dgamdalpha}
\end{figure}
In the decays $(D_1,D_2^*)\to D^*\pi_1\to D\pi_1\pi_2$, the angle between the
two
pions contains information about the initial spin state.
Let $\alpha$ be the angle between the momenta ${\bf p}_{\pi1}$ and ${\bf
p}_{\pi2}$, as measured in the rest frame of the excited meson.  Since the
$D_1$
and $D_2^*$ are separated by approximately $40\,$MeV and have intrinsic widths
of
$20-30\,$MeV, they overlap considerably.  The distribution in $\cos\alpha$ is a
function of where the pion's energy places it in relation to the two
resonances.
The formalism of Ref.~\cite{FP} may be used to extend the results of
Ref.~\cite{IW} to the case where the finite widths $\Gamma_{D_1}$ and
$\Gamma_{D_2^*}$ are taken into account.  We obtain
\begin{equation}\label{alphadist}
   {{\rm d}\Gamma\over{\rm d}\cos\alpha}\propto
   \left.{{\rm d}\Gamma\over{\rm d}\cos\alpha}\right|_{D_2^*}
   +\left.{{\rm d}\Gamma\over{\rm d}\cos\alpha}\right|_{D_1}\,,
\end{equation}
where
\begin{eqnarray}
   \left.{{\rm d}\Gamma\over{\rm d}\cos\alpha}\right|_{D_2^*}=&&
   {3\sin^2\alpha\over(E_\pi-\Delta_{21})^2+\Gamma^2_{D_2^*}/4}\,,\\
   \left.{{\rm d}\Gamma\over{\rm d}\cos\alpha}\right|_{D_1}=
   &&{{\cal A}_{d1}^2(1+3\cos^2\alpha)
   -{\cal A}_s{\cal A}_{d1}2\sqrt2(1-3\cos^2\alpha)
   \cos\eta\sin\psi+2{\cal A}_s^2\sin^2\psi\over
   (E_\pi-\Delta_{11})^2+\Gamma^2_{D_1}/4}\,.\nonumber
\end{eqnarray}
Here $\Delta_{21}$ and $\Delta_{11}$ are the resonant pion
energies (averaged over charge states, $\Delta_{21}=417\,$MeV and
$\Delta_{11}=383\,$MeV), and
\begin{eqnarray}
   {\cal A}_s&=&{\sqrt3\over2}\,{f\over(h/\Lambda_\chi)}
   \,{E_\pi\over E_\pi^2-m_\pi^2}\,,\nonumber\\
   {\cal A}_{d1} &=& 1+{8h_1\over3m_ch}
\end{eqnarray}
give the relative strength of the $s$-wave and $d$-wave transitions.
Since ${\cal A}_s$ contains a potentially large phase space enhancement, we
keep
${\cal A}_s\sin\psi$ in our expressions.  The two terms in
Eq.~(\ref{alphadist}) are normalized correctly with respect to each
other, but the overall normalization is arbitrary.  Note that while the
resonances overlap, they do not interfere in this distribution, a feature which
follows directly from Lorentz invariance.
In Fig.~\ref{dgamdalpha} we show the
distribution in $\cos\alpha$ for $E_\pi=(380, 400, 420)\,$MeV, for several
scenarios of $D_1$ decay:  pure $d$-wave (${\cal A}_s=0$ and $\psi=0$), pure
$s$-wave (${\cal A}_{d1}=0$ and $\psi=\pi/2$), and a mixed case with $h_1=0$
and $\psi=9^\circ$, and $|f|=0.85$.  This latter case (actually, two cases,
with $f=\pm0.85$) corresponds to the situation in which there is no enhancement
of the $d$-wave width, and the $s$-wave width is adjusted to give the correct
total width of the $D_1$.  We have also chosen a value of $\psi$ taken from the
quark
model~\cite{GIK}, and have set $\eta=0$.  We see that it should be possible to
distinguish between these various scenarios.

\subsubsection{One-pion distributions}

We may also consider another distribution which is tied more closely to the
fragmentation process by which the excited heavy meson is initially produced.
In the decay $(D_1,D_2^*)\to D^*\pi$, let $\theta$ be the angle between the
momentum of the pion and the fragmentation axis, as measured in the excited
meson rest frame.  In this frame, the fragmentation axis points back to the
hard
event in which the heavy quark was initially produced.  The angular
distribution in $\cos\theta$ depends not only on the
quantities $\{h,h_1,f,\eta,\psi\}$, but on the helicity distribution parameter
$w_{3/2}$~\cite{FP}.  This parameter describes the alignment with which the
light degrees of freedom of $J_\ell={3\over2}$ are produced in the creation of
the
$D_1$ or $D_2^*$.  The probabilities of the various helicity states along the
fragmentation axis are given by
\begin{eqnarray}
   P(J_\ell^3=\case3/2) &= P(J_\ell^3=-\case3/2) &=
   \case1/2w_{3/2}\,,\nonumber\\
   P(J_\ell^3=\case1/2) &= P(J_\ell^3=-\case1/2) &= \case1/2(1-w_{3/2})\,.
\end{eqnarray}
The parameter $w_{3/2}$ is a nonperturbative parameter of QCD, which is
well defined only in the heavy quark limit.  In Ref.~\cite{FP}, data from
ARGUS~\cite{ARGUS} on the decay $D_2^*\to D\pi$ was used to set the 90\%
confidence level upper limit $w_{3/2}<0.24$.  Models based on perturbative QCD
have yielded an estimate $w_{3/2}\approx0.25$~\cite{CW,Yuan}.

The distribution in $\cos\theta$ is considerably more complicated when finite
width effects are included, in part, because the interference between the
$D_1$ and $D_2^*$ resonances may not be neglected.  (We note that when
$w_{3/2}$ is
extracted from the decay $D_2^*\to D\pi$, there are no
interference effects, since the decay $D_1\to D\pi$ is prohibited.)   After a
straightforward calculation, we find
\begin{equation}\label{thetadist}
   {{\rm d}\Gamma\over{\rm d}\cos\theta} \propto
   \left.{{\rm d}\Gamma\over{\rm d}\cos\theta}\right|_{D_2^*}
   +\left.{{\rm d}\Gamma\over{\rm d}\cos\theta}\right|_{D_1}
   +\left.{{\rm d}\Gamma\over{\rm d}\cos\theta}\right|_{D_1-D_2^*}\,,
\end{equation}
where
\begin{eqnarray}
 \left.{{\rm d}\Gamma\over{\rm d}\cos\theta}\right|_{D_2^*}&=&
   {3(1+\cos^2\theta)+2w_{3/2}(1-3\cos^2\theta)
   \over(E_\pi-\Delta_{21})^2+\Gamma^2_{D_2^*}/4}\,,\\
 \left.{{\rm d}\Gamma\over{\rm d}\cos\theta}\right|_{D_1}&=&
   {1\over(E_\pi-\Delta_{11})^2+\Gamma^2_{D_1}/4}\left\{
   {\cal A}_{d1}^2\left[3(1+\cos^2\theta)+2w_{3/2}(1-3\cos^2\theta)\right]
   \right.\nonumber\\
   &&\qquad\left.\mbox{}+{\cal A}_s{\cal A}_{d1}
   2\sqrt2\sin\psi\cos\eta(-1+2w_{3/2})(1-3\cos^2\theta)
   +4{\cal A}_s^2\sin^2\psi\right\}\nonumber\\
 \left.{{\rm d}\Gamma\over{\rm d}\cos\theta}\right|_{D_1-D_2^*}&=&
   {(-1+2w_{3/2})(1-3\cos^2\theta)\over
   2[(E_\pi-\Delta_{21})^2+\Gamma^2_{D_2^*}/4]\,
   [(E_\pi-\Delta_{11})^2+\Gamma^2_{D_1}/4]}\\
   &&\quad\times\left\{\left({\cal A}_{d1}-{\cal A}_s\sqrt2\sin\psi\cos\eta
   \right)\left[4(E_\pi-\Delta_{11})(E_\pi-\Delta_{21})
   +\Gamma_{D_1}\Gamma_{D_2^*}\right]\right.\nonumber\\
   &&\qquad\left.\left.\mbox{}-{\cal A}_s2\sqrt2\sin\psi\sin\eta
   \left(E_\pi\Gamma_{D_1}-E_\pi\Gamma_{D_2^*}+\Delta_{11}\Gamma_{D_2^*}
   -\Delta_{21}\Gamma_{D_1}\right)\right]\right\}\nonumber\,.
\end{eqnarray}
Once again, the three terms in Eq.~(\ref{thetadist}) are normalized only with
respect to each other.
\begin{figure}
\epsfxsize=16cm
\hfil\epsfbox{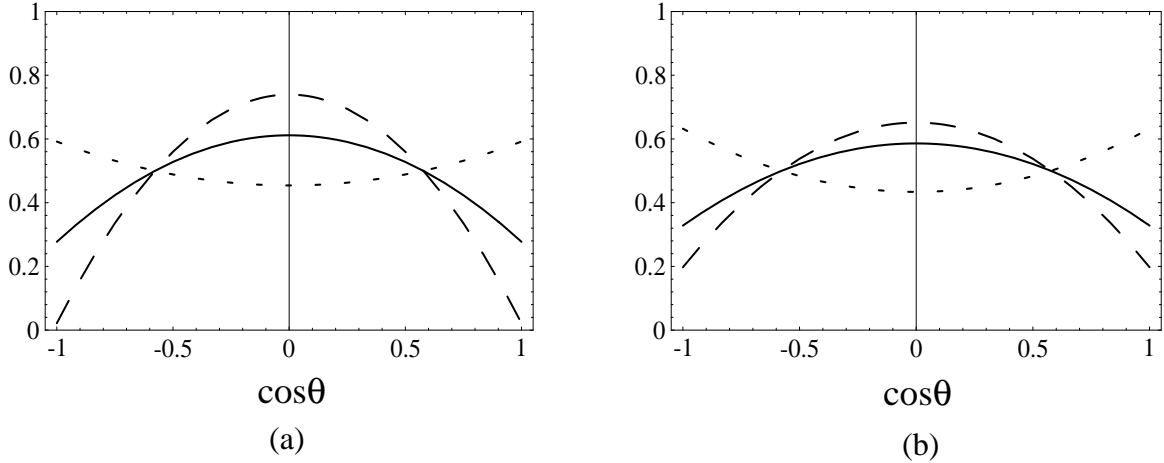}\hfill
\caption{The differential distribution in $\cos\theta$, (a) without, and (b)
with the finite widths of the $D_1$ and $D_2^*$.  We show curves for
$E_\pi=(380, 400, 420)\,$MeV, identified as in Fig.~1, and have set
$\psi=9^\circ$, $f=-0.85$, $h_1=w_{3/2}=\eta=0$.}
\label{widthcompare}
\end{figure}

To explore the importance of including the finite widths of the excited
resonances, in Fig.~\ref{widthcompare} we compare the distribution in
$\cos\theta$ with (a) $\Gamma(D_1)=\Gamma(D_2^*)=0$ and (b)
$\Gamma(D_1)=22\,$MeV, $\Gamma(D_2^*)=28\,$MeV, and scanning over
$E_\pi=(380, 400, 420)\,$MeV.  For the purpose of illustration, we set
$w_{3/2}=0$
and choose the parameters $h_1=0$, $\psi=9^\circ$ and $f=-0.85$ for the $D_1$
decay.

\begin{figure}
\epsfxsize=16cm
\hfil\epsfbox{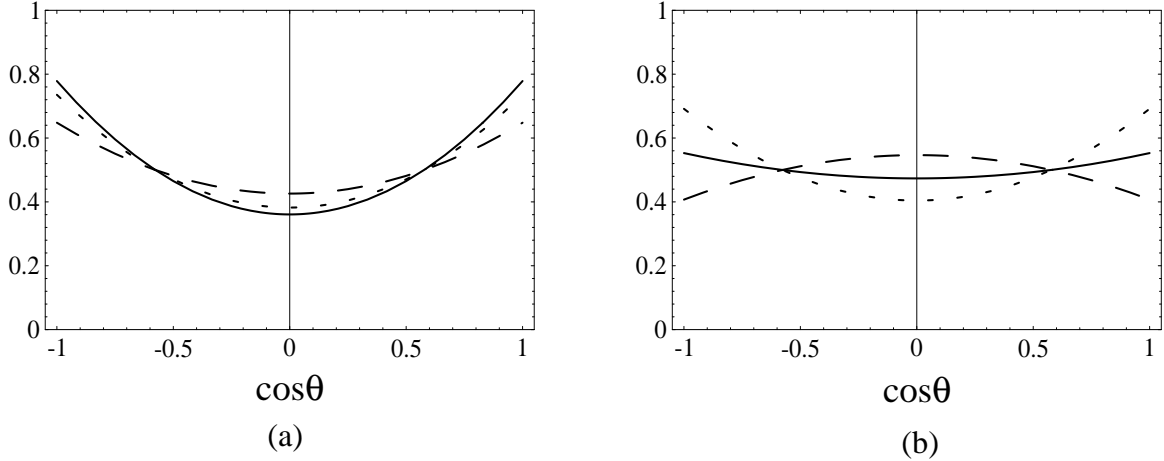}\hfill
\caption{The differential distribution in $\cos\theta$, for $w_{3/2}=0$, in the
case of (a) pure $d$-wave $D_1$ decay, and (b) pure $s$-wave $D_1$ decay.  We
have set $f=-0.17$, $\eta=0$, and $E_\pi=(380, 400, 420)\,$MeV.}
\label{wiszero}
\end{figure}

\begin{figure}
\epsfxsize=16cm
\hfil\epsfbox{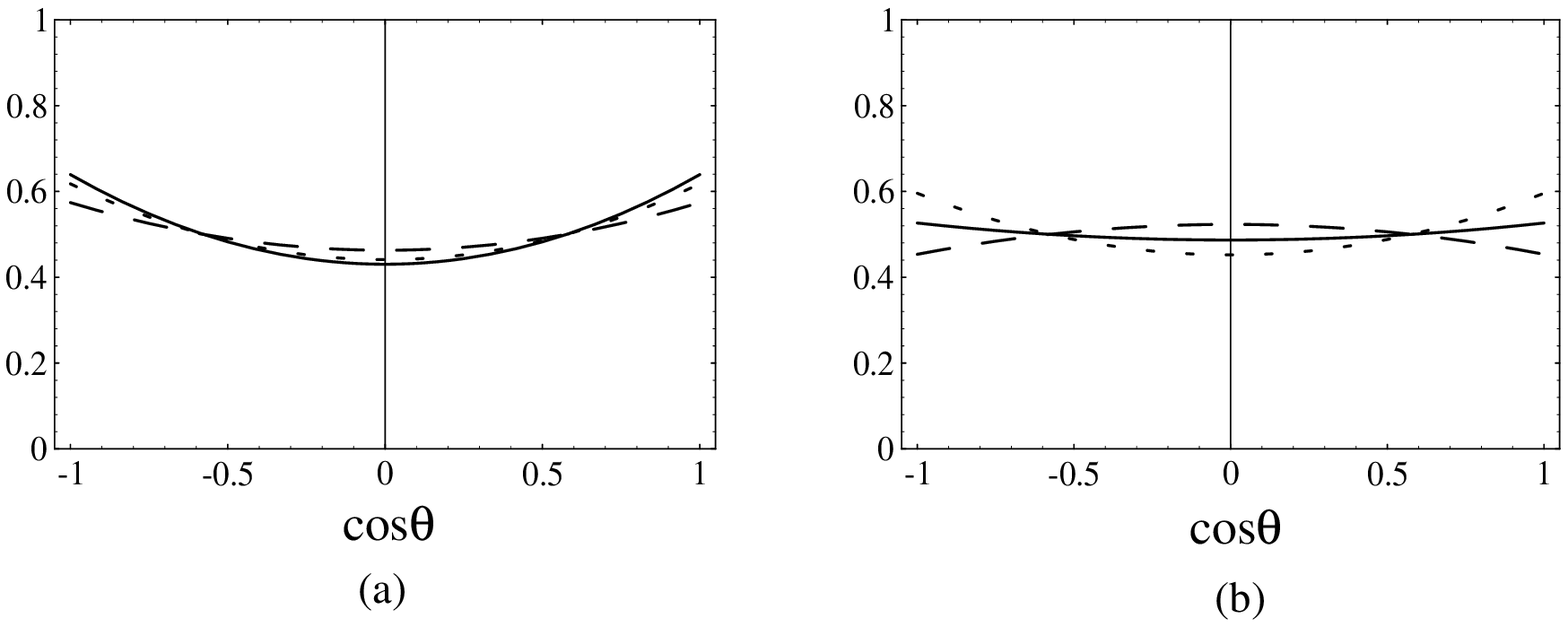}\hfill
\caption{The differential distribution in $\cos\theta$, for $w_{3/2}=0.25$, in
the case of (a) pure $d$-wave $D_1$ decay, and (b) pure $s$-wave $D_1$ decay.
We have set $f=-0.17$, $\eta=0$, and $E_\pi=(380, 400, 420)\,$MeV.}
\label{wisquarter}
\end{figure}

\begin{figure}
\epsfxsize=16cm
\hfil\epsfbox{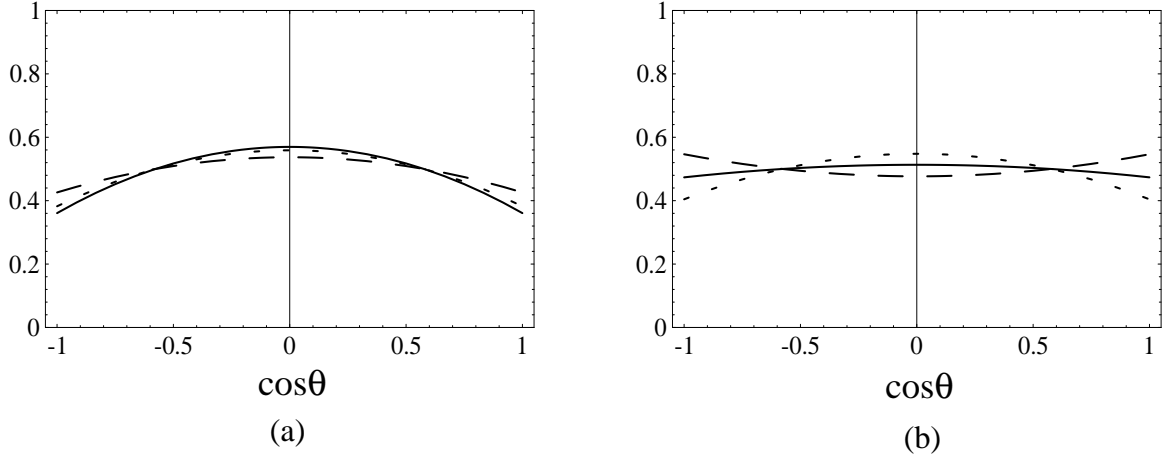}\hfill
\caption{The differential distribution in $\cos\theta$, for $w_{3/2}=0.75$, in
the case of (a) pure $d$-wave $D_1$ decay, and (b) pure $s$-wave $D_1$ decay.
We have set $f=-0.17$, $\eta=0$, and $E_\pi=(380, 400, 420)\,$MeV.}
\label{wis3quarter}
\end{figure}

\begin{figure}
\epsfxsize=16cm
\hfil\epsfbox{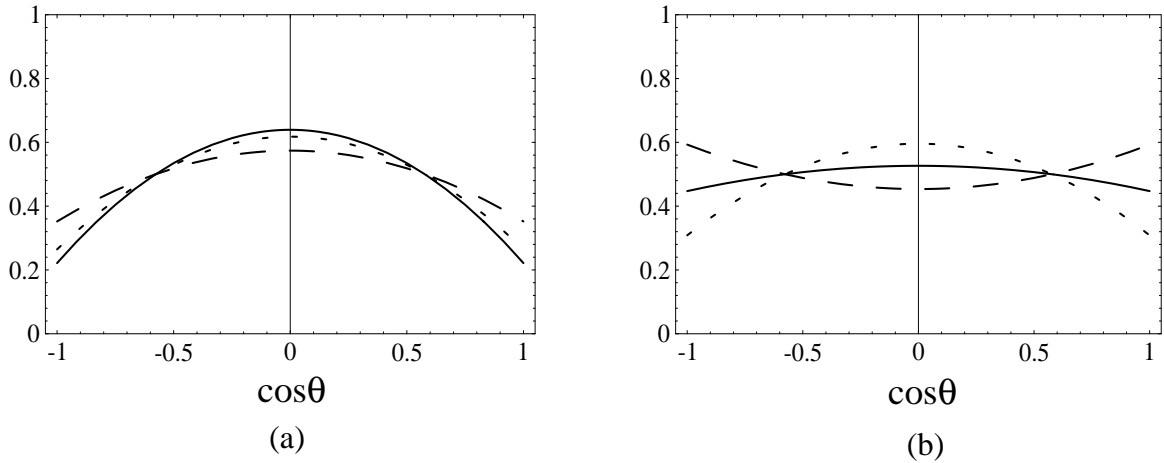}\hfill
\caption{The differential distribution in $\cos\theta$, for $w_{3/2}=1.0$, in
the case of (a) pure $d$-wave $D_1$ decay, and (b) pure $s$-wave $D_1$ decay.
We have set $f=-0.17$, $\eta=0$, and $E_\pi=(380, 400, 420)\,$MeV.}
\label{wisone}
\end{figure}

In Fig.~\ref{wiszero}, we set $w_{3/2}=0$ and compare pure $d$-wave and pure
$s$-wave
$D_1$ decays, for $f=-0.17$ and $E_\pi=(380, 400, 420)\,$MeV.  In
Fig.~\ref{wisquarter} we do
the same for $w_{3/2}=0.25$, in Fig.~\ref{wis3quarter} for $w_{3/2}=0.75$, and
in Fig.~\ref{wisone} for $w_{3/2}=1.0$.  (For $w_{3/2}=0.5$, all the
distributions are flat.)  Together, Figs.~\ref{wiszero}--\ref{wisone} give some
idea of the sensitivity of the distribution in $\cos\theta$ to the various
parameters describing the decay.  Note that the sensitivity to $w_{3/2}$ is
considerably enhanced if the $D_1$ decays via $d$-wave emission rather than
$s$-wave emission.

\section{Summary and Outlook}

Our detailed analysis of the leading effects of heavy spin symmetry violation
on the properties of excited charmed and bottom mesons has led to a number of
interesting results.  In particular, the width of the $D_1$, previously thought
to be anomalously large, is seen actually to be of a natural size.  Our
predictions for the properties of excited bottom mesons agree well with the
minimal data which exist so far, and will be tested soon in more detail.  We
presented detailed angular distributions for strong decays of excited charmed
mesons, which will eventually provide a more stringent test of the predictions
of our formalism.

Excited heavy mesons are important both for their own sake and for the
insight they give into the heavy quark expansion.  Since the most accurate
determinations of the CKM matrix element $V_{cb}$ involve the theoretical
application of HQET, it is crucial to understand how well the
$m_c,m_b\to\infty$ limit approximates the real world.  It is a matter of more
than academic interest whether the large width of the $D_1$ can be explained
naturally {\it within\/} the heavy quark expansion, because the answer to this
question affects our willingness to trust that the charm quark may be treated
as heavy in other contexts.  Similarly, it is worthwhile to do one's best to
extract parameters such as $w_{3/2}$, $f$ and $\psi$ from pion angular
distributions in strong decays.  Doing so, we learn not only about excited
heavy mesons themselves, but about whether one can indeed explain their
properties consistently in the context of the heavy quark expansion.

\acknowledgements
We thank Sam Osofsky for useful conversations.  This work was supported by  the
National Science Foundation under Grant No.~PHY-9404057.  A.F.~also
acknowledges
the National Science Foundation for National Young
Investigator Award No.~PHY-9457916, and the Department of
Energy for Outstanding Junior Investigator Award No.~DE-FG02-94ER40869.


\begin{references}

\bibitem{OPAL} R.~Akers {\it et al.}~(OPAL Collaboration), Z.\ Phys.\ {\bf
C66},
19 (1995); see also P.~Abreu {\it et al.}~(DELPHI Collaboration), Phys.\ Lett.\
{\bf B345}, 598 (1995).

\bibitem{CLEO} P.~Avery {\it et al.}~(CLEO Collaboration), Phys.\ Lett.\ {\bf
B331}, 236 (1994), Erratum, {\it ibid} {\bf B342}, 453 (1995); T.~Bergfeld {\it
et al.}~(CLEO Collaboration), Phys.\ Lett.\ {\bf B340}, 194 (1994); Y.Kubota
{\it
et al.} (CLEO Collaboration), Phys.\ Rev.\ Lett.\ {\bf 72}, 1972 (1994);
J.~Alexander {\it et
al.} (CLEO Collaboration), Phys.\ Lett.\ {\bf B303}, 377 (1993).

\bibitem{HQET} N.~Isgur and M.B.~Wise, Phys.\ Lett.\ {\bf B232}, 133 (1989);
Phys.\ Lett.\ {\bf B237}, 527 (1990).

\bibitem{PDG} Particle Data Group, Phys.\ Rev.\ {\bf D50}, 1173 (1994).

\bibitem{IW} N.~Isgur and M.B.~Wise, Phys.\ Rev.\ Lett. {\bf D66}, 1130 (1991);
M.~Lu, M.B.~Wise and N.~Isgur, Phys.\ Rev.\ {\bf D45}, 1553 (1992).

\bibitem{CT} P.~Cho and S.P.~Trivedi, Phys.\ Rev.\ {\bf D50}, 381 (1994).

\bibitem{EHQ} E.J.~Eichten, C.T.~Hill and C.~Quigg, Phys.\ Rev.\ Lett.\ {\bf
71},
4116 (1993).

\bibitem{FL} A.F.~Falk and M.~Luke, Phys.\ Lett. {\bf B292}, 119 (1992).

\bibitem{HHCPT} M.B.~Wise, Phys.\ Rev.\ {\bf D45}, 2188 (1992); G.~Burdman and
J.F.~Donoghue, Phys.\ Lett.\ {\bf B280}, 287 (1992); T.M.~Yan {\it et al.},
Phys.\ Rev.\ {\bf D46}, 1148 (1992).

\bibitem{FGGW} A.F.~Falk, H.~Georgi, B.~Grinstein and M.B.~Wise, Nucl.\ Phys.\
{\bf B343}, 1 (1990); A.F.~Falk, Nucl.\ Phys.\ {\bf B378}, 79 (1992).

\bibitem{LM} M.~Luke and A.V.~Manohar, Phys.\ Lett.\ {\bf B286}, 348 (1992).

\bibitem{FGL} H.~Georgi, Phys.\ Lett. {\bf B240}, 447 (1990); A.F.~Falk,
B.~Grinstein and M.~Luke, Nucl.\ Phys.\ {\bf B357}, 185 (1991).

\bibitem{TPC} J.C.~Anjos {\it et al.}~(Tagged Photon Spectrometer
Collaboration),
Phys.\ Rev.\ Lett.\ {\bf 62}, 1717 (1989).

\bibitem{GIK} S.~Godfrey and N.~Isgur, Phys.\ Rev.\ {\bf D32}, 189 (1985);
S.~Godfrey and R.~Kokoski, Phys.\ Rev.\ {\bf D43}, 1679 (1991).

\bibitem{FP} A.F.~Falk and M.E.~Peskin, Phys.\ Rev.\ {\bf D49}, 3320 (1994).

\bibitem{ARGUS} H.~Albrecht {\it et al.}~(ARGUS Collaboration), Phys.\ Lett.\
{\bf B221}, 422 (1989).

\bibitem{CW} Y.C.~Chen and M.B.~Wise, Phys.\ Rev.\ {\bf D50}, 4706 (1994).

\bibitem{Yuan} T.C.~Yuan, Phys.\ Rev.\ {\bf D51}, 4830 (1995).

\end{references}
\end{document}